\documentclass[prl,preprint,groupedaddress,showpacs]{revtex4}

\usepackage{amsmath}
\usepackage{amssymb}
\usepackage{amsfonts}
\usepackage{mathrsfs}
\usepackage{epsfig}
\usepackage{bm}

\newcommand{\ped}[1]{_{\text{#1}}}

\newcommand{\diff}[1]{\text{d}#1}
\newcommand{\pped}[1]{_{\scriptscriptstyle #1}}

\begin{document}

\title{Statistical kinetic treatment of relativistic binary collisions}

\author{F. Peano}\email{fabio.peano@ist.utl.pt}\author{M. Marti}\author{L. O. Silva}\email{luis.silva@ist.utl.pt}
\affiliation{GoLP/Instituto de Plasmas e Fus\~ao Nuclear, Instituto Superior T\'ecnico, 1049-001 Lisboa, Portugal}
\author{G. Coppa}
\affiliation{Dipartimento di Energetica, Politecnico di Torino, 10129 Torino, Italy}

\begin{abstract}
In particle-based algorithms, the effect of binary collisions is commonly described in a statistical way, using Monte Carlo techniques. It is shown that, in the relativistic regime, stringent constraints should be considered on the sampling of particle pairs for collision, which are critical to ensure physically meaningful results, and that nonrelativistic sampling criteria (e.g., uniform random pairing) yield qualitatively wrong results, including equilibrium distributions that differ from the theoretical J\"uttner distribution. A general procedure for relativistically consistent algorithms is provided, and verified with three-dimensional Monte Carlo simulations, thus opening the way to the numerical exploration of the statistical properties of collisional relativistic systems.
\end{abstract}

\date{\today}


\maketitle

The computer-assisted kinetic analysis of the behavior of many-particle systems is fundamental in several areas of modern physics, ranging from astrophysics (evolution of cosmological systems, dark-matter dynamics) \cite{Astro} to the physics of space and laboratory plasmas (relativistic shocks, spacecraft shielding, laser- and plasma-based particle acceleration, inertial confinement fusion) \cite{osiris}. 
When the effect of close encounters (collisions) can be neglected, and particles can be assumed to interact via smoothly varying, long-range forces, kinetic particle-mesh algorithms \cite{Hockney} are effective and versatile tools to study the evolution of the phase-space distribution function of each particle species in the system. Important examples are the particle-in-cell (PIC) method \cite{Hockney,Birdsall,osiris}, which provides a self-consistent description of the kinetics of collisionless plasmas over distances much longer than the Debye screening length (as described by the Vlasov-Maxwell set of equations), and hybrid methods, mixing kinetic and fluid approaches \cite{Gargate,Sherlock}. However, in situations where the inclusion of collisional effects in the model is critical, or when dealing with collisional-dominated many-body systems, the collision processes must be dealt with using physically consistent algorithms, in order to provide a correct description of the relevant statistical properties. 

A common way to include the effect of binary collisions in particle-based algorithms (cf. Ref. \cite{Takizuka,Vahedi}) is by locally changing the momenta of a suitable statistical sample of particle pairs using Monte Carlo (MC) techniques \cite{Bird_book}. This approach, often referred to as Direct Simulation Monte Carlo (DSMC) method, provides an accurate solution of the Bolzmann equation \cite{Bird_book,Bird,Wagner,Cercignani} (which is valid for dilute systems), and has been successfully employed in molecular gas dynamics \cite{Bird_book,Gallis} and plasma physics \cite{Sherlock,Takizuka,Vahedi,Nanbu,Miller,Larson,Kawamura,Birdsall_IEEE,Wilson}, mostly in the nonrelativistic regime. 
The application of these techniques to situations where the particle velocities are relativistic is relevant to many scenarios in High-Energy-Density Science, such as fast ignition of fusion targets (cf. Ref. \cite{Sentoku}), fast electron transport in solid targets, proton acceleration, or shocks.
As discussed in this Letter, the extension to relativistic regimes can not be achieved merely by guaranteeing energy-momentum conservation. Indeed, special relativity imposes further constraints on the way particle pairs are chosen for collision, 
independently of the particular type of collision process considered, even when the microscopic dynamics of each collision is modeled correctly. Overlooking these constraints on pair selection leads to unphysical results, with consequences as extreme as the systematic appearance of qualitatively wrong equilibrium distribution functions and energy-temperature relations (cf. Refs. \cite{Lehmann,Dunkel} and discussions in Refs. \cite{Cubero,Chacon}).

\begin{figure}[!htb]
\centering \epsfig{file=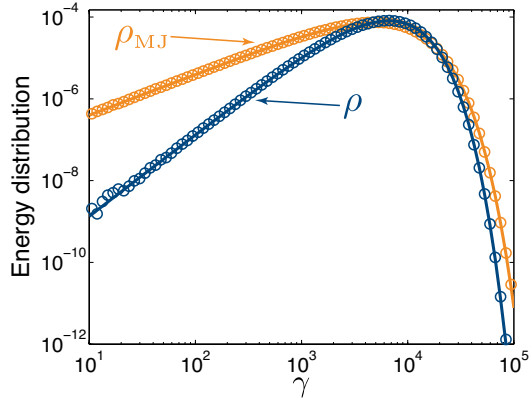, width=2.75in}
\caption{(Color online) Equilibrium energy spectra $\rho\ped{eq}(\gamma)$ obtained using the relativistically consistent law $\mathcal{A}\left({\bf v}_a,{\bf v}_b\right)\propto (1-{\bf v}_a \cdot {\bf v}_b)$ (dark) and the nonrelativistic approximation $\mathcal{A}\left({\bf v}_a,{\bf v}_b\right) = \mathrm{Constant}$ (light). Markers: simulation results; solid lines: $\rho(\gamma)$ and $\rho\pped{\text{MJ}} (\gamma)$ as in the text. The numerical data have been taken at a single time instant after approximately $10-20$ collisions per particle have occurred. Units are normalized so that $\int\!\rho\ped{eq}(\gamma)\diff{\gamma} = 1$.}
\label{fig:JMJ}
\end{figure}

In this Letter, the general procedure for the statistical kinetic treatment of binary collisions in the relativistic regime is described, thus providing a consistent framework for the exploration of relativistic many-particle systems with MC simulations.
Results from three-dimensional (3D) ultrarelativistic MC simulations with $\gtrsim 10^8$ computational particles are presented, illustrating the technique and reproducing the correct equilibrium distribution function over several orders of magnitudes in energy and particle number (Fig. \ref{fig:JMJ}). A systematic origin of conflicting results \cite{Lehmann,Dunkel,Cubero,Chacon} is identified, and, within the present kinetic framework, a simple interpretation of the numerical results published in a recent Letter by Cubero {\it et al.} \cite{Cubero} is given.

In relativistic kinetic theory \cite{DeGroot,Cercignani_2,Chacon}, the number of collisions, $\Delta N$, occurring within the space-time element $\Delta {\bf x}\Delta t$ about $\left({\bf x},t\right)$ between particles $a$, having momenta in the range $({\bf p}_a,{\bf p}_a+\Delta {\bf p}_a)$, and particles $b$, having momenta in the range $({\bf p}_b,{\bf p}_b+\Delta {\bf p}_b)$, is 
\begin{equation}
\Delta N \!\!=\! \mathcal{A} \! \left({\bf v}_a,\!{\bf v}_b\right) \!f_a \! \left({\bf x},\!{\bf p}_a,\!t\right) \!f_b \! \left({\bf x},\!{\bf p}_b,\!t\right) \!\Delta {\bf p}_a \Delta {\bf p}_b\Delta {\bf x}\Delta t \text{,}
\label{eq:dN}
\end{equation}
where $f_a$ and $f_b$ are the distribution functions of species $a$ and $b$ (assumed to be smooth enough to neglect differences in the space-time coordinates before and after collisions \cite{DeGroot}), and where $\mathcal{A}\left({\bf v}_a,{\bf v}_b\right)$ determines the collision probability as a function of the velocities ${\bf v}_a$ and ${\bf v}_b$. 
Since $\Delta N$ is a relativistic invariant, and so are $f_a$, $f_b$, and $\Delta {\bf x}\Delta t$, then $\mathcal{A}\left({\bf v}_a,{\bf v}_b\right) \Delta {\bf p}_a \Delta {\bf p}_b$, and hence $\mathcal{A}\left({\bf v}_a,{\bf v}_b\right)\gamma_a\gamma_b$, must be invariant as well \cite{Landau}, with $\gamma_{a,b}=(1-v_{a,b}^2)^{-1/2}$ (a system of units  where the speed of light is unitary is adopted). Introducing the total cross section $\sigma(v\ped{r})$ yields $\mathcal{A}\left({\bf v}_a,{\bf v}_b\right)\gamma_a\gamma_b=v\ped{r}\sigma(v\ped{r})(1-v\ped{r}^2)^{-1/2}$, which leads to the general expression
\begin{equation}
\mathcal{A}\left({\bf v}_a,{\bf v}_b\right)= v\ped{r}\sigma(v\ped{r})(1-{\bf v}_a \cdot {\bf v}_b) \text{,}
\label{eq:A}
\end{equation}
where $v\ped{r}=[({\bf v}_a-{\bf v}_b)^2-({\bf v}_a\times{\bf v}_b)^2]^{1/2}/(1-{\bf v}_a \cdot {\bf v}_b)$ is the absolute value of the relative velocity in a reference frame where one particle is at rest \cite{Chacon,DeGroot,Landau}.

One important consequence of Eq. \eqref{eq:A} is that $\mathcal{A}\left({\bf v}_a,{\bf v}_b\right)$ cannot be a function of a single invariant parameter (e.g., $v\ped{r}$), as it is in the nonrelativistic regime, where $\mathcal{A}\ped{NR}\left({\bf v}_a,{\bf v}_b\right)=\sigma(|{\bf v}_a-{\bf v}_b|)|{\bf v}_a-{\bf v}_b|$.  
Any violation of this essential constraint in calculations or simulations breaks the invariance of $\Delta N$, leading to qualitatively unphysical results, notably to equilibrium distribution functions differing from the stationary solution of the Boltzmann equation, which, for relativistic systems, is the J\"uttner function $f\pped{\text{J}}\left({\bf x},{\bf p}\right)\propto\exp\{\Gamma_{\bf U}[{\bf U}\cdot{\bf p}-\epsilon_0\gamma({\bf p})]/k_BT\}$ \cite{Juttner,DeGroot,Cercignani_2,Chacon,Synge}, where $\epsilon_0$ is the rest energy, the constant ${\bf U}$ is the equilibrium mean velocity \cite{DeGroot}, $\Gamma_{\bf U}=\sqrt{1-{\bf U}^2}$, $k_B$ is the Boltzmann constant, and the invariant constant $T$ is the equilibrium temperature measured in the reference frame where ${\bf U}={\bf 0}$.

In the statistical treatment of relativistic binary collisions, it is mandatory to adopt a procedure that satisfies not only the fundamental conservation laws, such as the conservation of the total 4-momentum, but also the relativistic invariance of $\Delta N$, a subtler but equally important requirement. 
In the nonrelativistic regime, this is usually not a concern, because the Galilean invariance of $\Delta N$ is trivially satisfied, with all quantities in Eq. \eqref{eq:dN} being invariant.
Thus, particular attention is needed whenever applying nonrelativistic approximations, since these may violate the invariance of $\Delta N$: 
a striking, paradigmatic example is the assumption of a uniform collision probability, $\mathcal{A}\left({\bf v}_a,{\bf v}_b\right) = \mathrm{Constant}$, corresponding to random pairing in MC algorithms, as commonly employed in nonrelativistic or weakly relativistic PIC simulations \cite{Takizuka,Nanbu,Sentoku}. According to special relativity, such an assumption is unphysical, leading to a wrong equilibrium distribution, described by a modified J\"uttner function, 
$f\pped{\text{MJ}}\left({\bf x},{\bf p}\right)\propto\exp\{\Gamma_{\bf U}[{\bf U}\cdot{\bf p}-\epsilon_0\gamma({\bf p})]/k_BT\pped{\text{MJ}}\}/\{\Gamma_{\bf U}[\epsilon_0\gamma({\bf p})-{\bf U}\cdot{\bf p}]\}$, and, contextually, to a wrong equilibrium temperature, $T\pped{\text{MJ}}$.  In the recent literature, $f\pped{\text{MJ}}$ has been proposed as a plausible extension of the nonrelativistic Maxwell-Boltzmann distribution to relativistic systems \cite{Lehmann,Dunkel}, but this possibility has been recently ruled out using one-dimensional (1D) numerical simulations \cite{Cubero}. As shown here, $f\pped{\text{MJ}}$ is obtained in MC algorithms whenever $\mathcal{A}\left({\bf v}_a,{\bf v}_b\right)$ is erroneously assumed to be a function of the single invariant parameter $v\ped{r}$, e.g. with uniform random pairing, independently of the particular choice of $\sigma(v\ped{r})$ [as in the nonrelativistic case, $\sigma(v\ped{r})$ merely affects relaxation processes, having no effect on the equilibrium distribution].

In order to obtain the correct physical results in particle-based kinetic algorithms, with an MC approach, it is sufficient to adopt a three-step procedure: given a collection of particles contained in a spatial region $\Delta {\bf x}$, the momenta of a statistical sample of particle pairs undergoing a given collision process are updated over a time interval $\Delta t$ by 
\begin{enumerate}
  \item Sampling the colliding pairs according to the relativistic expression of $\mathcal{A}\left({\bf v}_a,{\bf v}_b\right)$ given in Eq. \eqref{eq:A} (e.g., with standard rejection methods \cite{Bird_book}), so as to guarantee the invariance of $\Delta N$. Depending on the problem, the number of colliding pairs must be chosen appropriately, ensuring that, on average, the correct number of collisions is performed, and the correct collision frequency is recovered \cite{Bird_book,Bird,Vahedi}.   
  \item Deciding the output of each collision, using the differential cross section \cite{DeGroot,Cercignani_2,Chacon} to evaluate the scattering angle, so as to guarantee that the microscopic details of the collision process are modeled correctly. For inelastic collisions (e.g., reactions, ionizations, recombinations, pair creation/annihilation), this may involve particle generation and removal. 
  \item Updating the momenta of all particles resulting from each collision, obeying to the relevant conservation laws (e.g., the conservation of the total energy-momentum and of the total electric charge). As an example, for an elastic collision between particles $a$ and $b$, this step is conveniently performed by transforming ${\bf p}_a$ and ${\bf p}_b$ to the center-of-momentum frame, rotating the momenta by the appropriate scattering angle, and transforming the new momenta back to the laboratory frame \cite{Sentoku}. 
\end{enumerate}
This procedure provides a correct description of the collisional dynamics, as predicted by the Boltzmann equation \cite{Bird}, and correctly yields the equilibrium distribution function $f\pped{\text{J}}$, independently of the specific cross section, and for all energy ranges. 

As a test for the algorithm, the evolution to equilibrium of many-particle systems in conditions ranging from nonrelativistic to ultrarelativistic regimes has been investigated with massively parallel, 3D MC simulations based on the Osiris 2.0 framework \cite{osiris}, employing up to $10^9$ computational particles. In the example shown here, the equilibrium distribution of a single species of ultrarelativistic particles undergoing elastic, isotropic collisions is analyzed using $2 \times 10^8$ computational particles. 
A monoenergetic initial distribution has been used, $f({\bf x},{\bf p},t=0)\propto\delta[\gamma({\bf p})-\gamma_0]$ with $\gamma_0 = 10^{4}$ and mean velocity ${\bf U}={\bf 0}$, and $\sigma(v\ped{r})\propto1/v\ped{r}$ has been assumed, thus yielding $\mathcal{A}\left({\bf v}_a,{\bf v}_b\right) \propto (1-{\bf v}_a \cdot {\bf v}_b)$.

In order to provide a clear evidence that the equilibrium distribution $f\ped{eq}\left({\bf x},{\bf p}\right)$ accurately reproduces $f\pped{\text{J}}\left({\bf x},{\bf p}\right)$, the corresponding energy distribution  
\begin{equation}
\rho\ped{eq}(\gamma) = \iint \!\! \delta\!\left(\sqrt{\epsilon_0^2+{\bf p}^2}-\epsilon_0\gamma\right) f\ped{eq}\left({\bf x},{\bf p}\right)\diff{{\bf p}}\diff{{\bf x}}
\label{eq:rho}
\end{equation}
has been constructed directly from the numerical data (by counting the number of particles having energy within finite intervals on the $\gamma$ axis), and plotted over a wide range of $\gamma$, spanning several orders of magnitudes (Fig. \ref{fig:JMJ}). The simulated equilibrium energy distribution accurately reproduces the theoretical curve $\rho(\gamma)=\gamma\sqrt{\gamma-1}\exp\left(-\epsilon_0\gamma/k_BT\right)$, obtained by replacing $f\ped{eq}$ with $f\pped{\text{J}}$ in Eq. \eqref{eq:rho}, where the equilibrium temperature $k_BT=3.33 \times 10^3 \epsilon_0$ is calculated from the initial mean energy as $\left<\gamma\right> \approx 1+3k_BT/\epsilon_0$, the ultrarelativistic limit of the energy-temperature relation
\begin{equation}
\left<\gamma\right> = \dfrac{\displaystyle\iint \!\! \sqrt{\epsilon_0^2+{\bf p}^2} f\pped{\text{J}} \diff{{\bf p}}\diff{{\bf x}}}{\epsilon_0 \!\! \displaystyle\iint \!\! f\pped{\text{J}} \diff{{\bf p}}\diff{{\bf x}}} =
\frac{K_3\left(\frac{\epsilon_0}{k_BT}\right)}{K_2\left(\frac{\epsilon_0}{k_BT}\right)}-\frac{k_BT}{\epsilon_0} \text{,}
\label{eq:TJ}
\end{equation}
where $K_n$ denotes the $n$th order modified Bessel function of the second kind \cite{Abramovitz}. The numerical results are in complete agreement with the theoretical curve, correctly reproducing variations spanning eight orders of magnitudes in $\rho\ped{eq}$ (Fig. \ref{fig:JMJ}).
 
The shape of $\rho\ped{eq}(\gamma)$ obtained by (incorrectly) sampling the colliding pairs according to the nonrelativistic approximation $\mathcal{A}\left({\bf v}_a,{\bf v}_b\right) = \mathrm{Constant}$, is also shown. The distribution reproduces the modified curve $\rho\pped{\text{MJ}} (\gamma) = \sqrt{\gamma-1}\exp\left(-\epsilon_0\gamma/k_BT\pped{\text{MJ}}\right)$, obtained by replacing $f\ped{eq}$ with $f\pped{\text{MJ}}$ in Eq. \eqref{eq:rho}. The equilibrium temperature $k_BT\pped{\text{MJ}} = 5 \times 10^3 \epsilon_0$ is calculated from the initial mean energy as $\left<\gamma\right>\pped{\text{MJ}} \approx 1+2k_BT\pped{\text{MJ}}/\epsilon_0$, which is the ultrarelativistic limit of the modified energy-temperature relation
\begin{equation}
\left<\gamma\right>\pped{\text{MJ}} = \dfrac{\displaystyle\iint \!\! \sqrt{\epsilon_0^2+{\bf p}^2} f\pped{\text{MJ}} \diff{{\bf p}}\diff{{\bf x}}}{\epsilon_0 \!\! \displaystyle\iint \!\! f\pped{\text{MJ}} \diff{{\bf p}}\diff{{\bf x}}} =
\frac{K_2\left(\frac{\epsilon_0}{k_BT\pped{\text{MJ}}}\right)}{K_1\left(\frac{\epsilon_0}{k_BT\pped{\text{MJ}}}\right)} \text{.}
\label{eq:TMJ}
\end{equation}
Although still complying with the energy-momentum conservation (step 3 above), this result is unphysical, because it violates the relativistic invariance of $\Delta N$: if performed within a Lorentz-boosted reference frame \cite{Vay}, with boost factor $\gamma\ped{b}$, the same simulation would exhibit an artificial increase of the total number of collisions by a factor $\gamma\ped{b}^2$, as can be readily verified from Eqs. \eqref{eq:dN} and \eqref{eq:A}, thus yielding a significantly different dynamical evolution of the system and a wrong equilibrium state.

The present analysis also allows for a straightforward kinetic interpretation of the numerical results recently presented in Ref. \cite{Cubero}, where molecular dynamics (MD) simulations of a 1D system composed of two species of colliding particles have been used to provide a numerical confirmation that the equilibrium one-particle distribution of a dilute relativistic gas is described by the J\"uttner fuction $f\pped{\text{J}}$, as opposed to the modified J\"uttner function, $f\pped{\text{MJ}}$.
The 1D system considered in Ref. \cite{Cubero} is a collection of impenetrable particles undergoing binary collisions, wherein interactions are zero-range and particles act as infinitely extended rigid sheets. Each collision is a localized event in space-time, with a fully deterministic outcome. Between collisions, particles are free-streaming, with the Hamiltonian of the system being the linear superposition of the relativistic Hamiltonians of each free particle, $\mathscr{H}_n({\bf x},{\bf p}) = \sqrt{\epsilon_0^2+{\bf p}_n^2}$ for the $n$th particle. This allows for a fully deterministic numerical solution of the equations of motion via standard MD techniques \cite{MD}.
In the kinetic approximation, the basic statistical properties of the system analyzed in Ref. \cite{Cubero} (i.e., the equilibrium one-particle distribution function integrated over space) can be investigated using the relativistic Boltzmann equation, whose stationary solution is $f\pped{\text{J}}$ \cite{DeGroot,Cercignani_2,Chacon,Synge}.
In 1D, Eq. \eqref{eq:dN} reduces to $\Delta N = \mathcal{P}(v\ped{r})|v_a-v_b| f_a\left(x,p_a,t\right) f_b\left(x,p_b,t\right) \Delta p_a \Delta p_b \Delta x\Delta t$, 
where $\mathcal{P}(v\ped{r})$ is the probability for an $a$-$b$ encounter to result in a collision, with the limit $\mathcal{P}(v\ped{r}) \rightarrow 1$ corresponding to impenetrable particles, as considered in Ref. \cite{Cubero}.
Determining each collision event exactly, the MD algorithm used in Ref. \cite{Cubero} implicitly guarantees the invariance of $\Delta N$, thus yielding the correct distribution function $f\pped{\text{J}}$.
Statistical approaches recover the same result, independently of the particular shape of $\mathcal{P}(v\ped{r})$, provided that colliding pairs are sampled according to the relativistically consistent law $\mathcal{P}(v\ped{r})|v_a-v_b|=v\ped{r}\mathcal{P}(v\ped{r})(1-v_a v_b)$. 
As in the 3D case (Fig. \ref{fig:JMJ}), if the colliding pairs are erroneously sampled according to a nonrelativistic, one-parameter law of the form $\mathcal{P}(v\ped{r})v\ped{r}$, the modified function $f\pped{\text{MJ}}$ is always obtained.
The formal proof is straightforward: in 1D, the collision integrals \cite{Landau} expressing the net change per unit time in the distribution function of particles $a$ and $b$ due to collisions read $J_{a,b} =\int \mathcal{P}(v\ped{r})|v_a-v_b|(f_a^\prime f_b^\prime-f_af_b)dp_{b,a}$, 
where $f_{a,b}^\prime=f_{a,b}(x,p_{a,b}^\prime,t)$, with $p_{a,b}^\prime$ denoting momenta after collisions. Setting the local entropy production $s(x,t) \propto -\sum_{\alpha=a,b} \int \log (f_{\alpha}) J_{\alpha} \diff{p_{\alpha}}$ \cite{DeGroot} to zero, then yields $f_a^\prime f_b^\prime=f_af_b$, leading, for both species, to the equilibrium distribution function $f\pped{\text{J}}$, with same equilibrium temperature $T$. If the calculation is repeated after replacing $\mathcal{P}(v\ped{r})|v_a-v_b|$ with the nonrelativistic law $\mathcal{P}(v\ped{r})v\ped{r}$, the collision integrals become $\tilde{J}_{a,b} =\int \mathcal{P}(v\ped{r})|v_a-v_b|(\gamma_a^\prime f_a^\prime \gamma_b^\prime f_b^\prime-\gamma_a f_a\gamma_b f_b)dp_{b,a}$, 
yielding $\gamma_a^\prime f_a^\prime \gamma_b^\prime f_b^\prime=\gamma_a f_a\gamma_b f_b$, which leads to the modified equilibrium distribution $f\pped{\text{MJ}}$, with a modified equilibrium temperature $T\pped{\text{MJ}}$. Hence, from a purely mathematical point of view, $f\pped{\text{MJ}}$ could be considered as the stationary solution of a modified relativistic Boltzmann equation (cf. Conclusions in Ref. \cite{Chacon} and references therein), with collision integrals of the form $\tilde{J}_{a,b}$. Again, such an equation would violate the relativistic invariance of $\Delta N$, thus being physically inconsistent.

In summary, the problem of providing a consistent statistical description of relativistic binary collisions in dilute many-particle systems has been analyzed using the standard relativistic kinetic theory, showing that rigorous constraints hold on the way particle pairs are chosen for collision, and that nonrelativistic approximations (such as a uniform collision probability) are forbidden. By breaking the relativistic invariance of the number of collision events in a given space-time region, these approximations lead to unphysical, conflicting results, notably modified equilibrium distribution functions. Thus, in any calculation or simulation based on statistical sampling of colliding particles, the invariance of $\Delta N$ constitutes a fundamental validity criterion, as important as the more obvious energy-momentum conservation, in order to guarantee that results are physically meaningful, with the equilibrium distribution being described by the J\"uttner function, as predicted by the relativistic Boltzmann equation.
The present discussion thus provides the framework for the detailed exploration, via Monte Carlo simulations, of the statistical properties of multi-dimensional, collisional systems in the relativistic regime.

\begin{acknowledgments}
This work was partially supported by FCT (Portugal) through grant POCI/FIS/66823/2006.
\end{acknowledgments}

\end{document}